# TEMPERATURE CHANGES OF SPECTRA OF THE LATTICE AND SURFACE OSCILLATIONS OF ORGANIC MOLECULAR NANO-CRYSTALS (COMPUTER MODELING)

M.A. Korshunov[*]

*L.V. Kirensky Institute of physics, Siberian branch of the Russian Academy of Sciences, 660036 Krasnoyarsk, Russia*

**ABSTRACT.** Dependence of frequency spectra of the lattice oscillations of organic nano-crystals on temperature is studied at presence of vacancies in structure. In a frequency spectrum a number of additional lines monotonously changing with temperature is observed. Also, the spectrum of the surface oscillations is calculated at temperature change. At the same time, the low-frequency line has non-linear behavior with temperature change.

At studying of organic molecular crystals at reduction of their sizes the role of a surface and the surface oscillations that should find the reflection in Raman spectra increases. In spectra, as finds the reflection and presence of flaws in particular vacancies. To part display in spectra of the surface oscillations from a spectrum of the lattice oscillations in the presence of vacancies in crystal structure, calculation of these spectra is carried out. To reveal display of those or other flaws it is possible, having carried out temperature examinations. Therefore, calculations of the lattice and surface oscillations have been carried out at temperature change. As object of examination the typical organic molecular crystal para-dichlorobenzol has been chosen. At calculations interaction between molecules paid off on a method atom-atom of potentials. Interaction coefficients got out such that the spectrum of the lattice oscillations of an ideal crystal coincided with the observational spectrum of small frequencies. Further at calculations of crystals with flaws they did not vary. Calculations were spent with use of a method the Dyne [1]. This method allows to carry out calculations of frequency spectra of disorder molecular crystals. On the basis of calculations histograms which show probability of display of lines of a spectrum in the chosen frequency interval have been gained. Change of temperature in a nano-crystal was modeled by change of parameters of a lattice which are taken from work [2] in which dependence of parameters of a lattice on temperature for para-dichlorobenzol is observationally gained.

At first calculations of the lattice oscillations have been carried out at presence in structure of vacancies. In Figs. 1-3 change of the histogram of frequencies of the lattice oscillations from temperature (Fig. 1 - 223К, Fig. 2 - 293K, and Fig. 3 - 303K) is shown. As we see there is the monotonous change of lines of a spectrum to temperature there are additional lines. At 293K in a spectrum occurrence of lines in the field of 80 cm$^{-1}$ unlike a spectrum of an ideal crystal is observed.

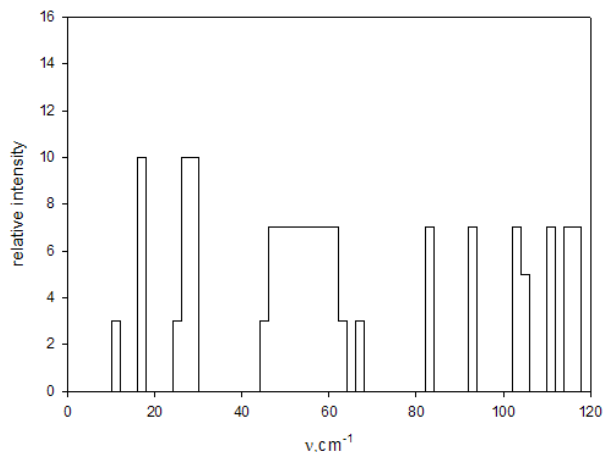

**Fig. 1.** The histogram of a frequency spectrum of para-dichlorobenzol with vacancies at temperature 223К.

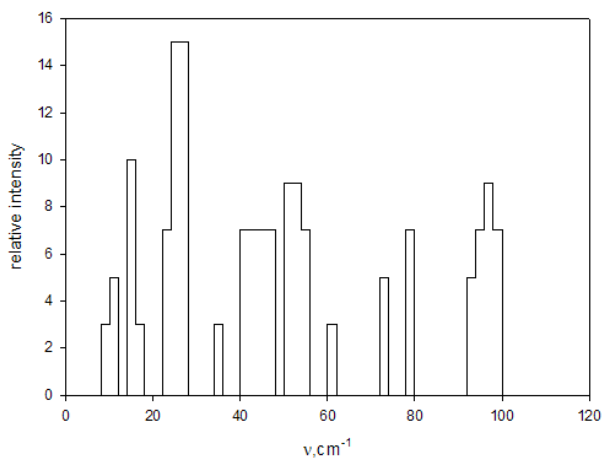

Fig. 2. The histogram of a frequency spectrum of para-dichlorobenzol with vacancies at temperature 293К.

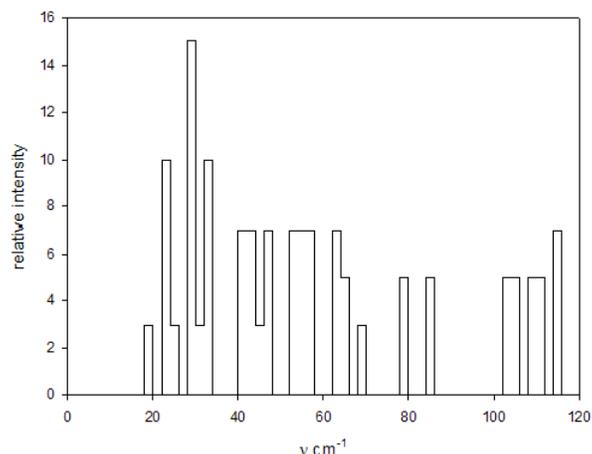

Fig. 4. The histogram of a frequency spectrum of the surface oscillations of para-dichlorobenzol at temperature 223К.

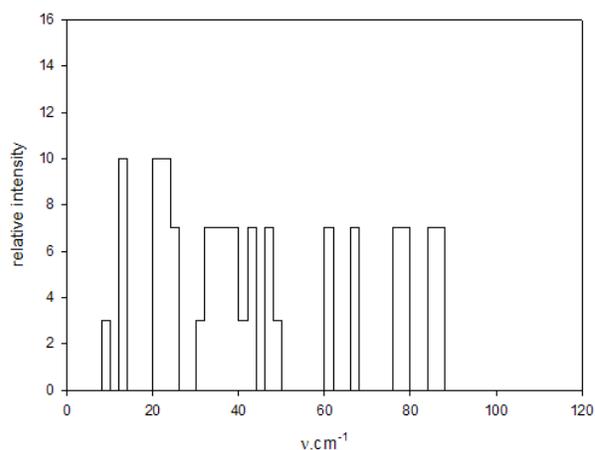

Fig. 3. The histogram of a frequency spectrum of para-dichlorobenzol with vacancies at temperature 303К.

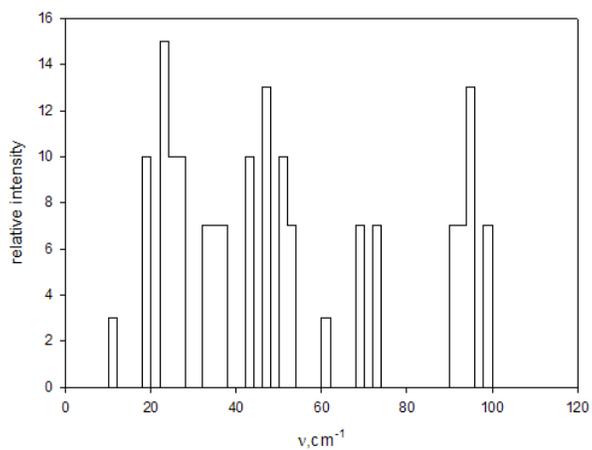

Fig. 5. The histogram of a frequency spectrum of the surface oscillations of para-dichlorobenzol at temperature 293К.

Histograms of the surface oscillations are given in Figs. 4-6. At the same temperatures as spectra of the lattice oscillations. If the observational spectrum from a nano-crystal that in a spectrum takes place there will be the surface oscillations from all facets of a crystal unlike crystals of the major size. Apparently in a spectrum there are additional lines, as well as in a case with vacancy. Thus the low-frequency line in the field of ~15 cm$^{-1}$ has non-linear behavior with temperature change. Other lines monotonously change with temperature. At temperature 293К, unlike spectra of the lattice oscillations of a crystal with vacancy it is not observed lines in the field of 80 cm$^{-1}$.

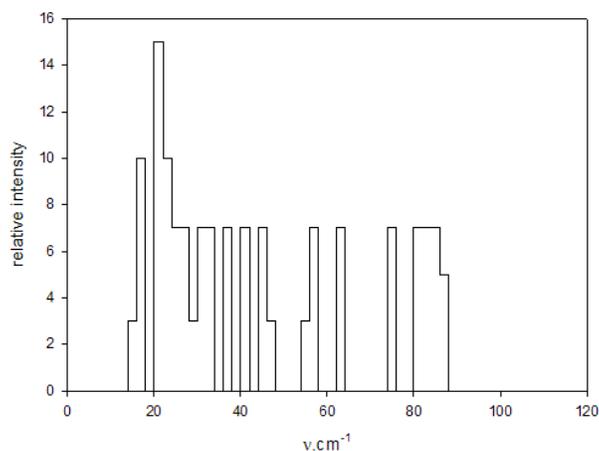

Fig. 6. The histogram of a frequency spectrum of the surface oscillations of para-dichlorobenzol at temperature 303К.

Thus differences observed in behavior of lines of a spectrum of the lattice oscillations of a nano-crystal with vacancy and the surface oscillations with temperature change allows to interpret the observational spectra of organic molecular nano-crystals.